\documentclass[runningheads]{llncs}

\usepackage{epsfig}
\usepackage{verbatim}
\usepackage{url}

\def\Oh{\mbox{\rm O}}

\def\reverse#1{\hat{#1}}

\def\SA{\mbox{\rm {\sf SA}}}

\def\rank{\textsf{rank}}

\def\X{\mathsf{X}}
\def\MS{\mbox{\rm {\sf MS}}}
\def\B{\mathsf{B}}

\def\A{\mathsf{A}}

\def\Y{\mbox{\rm {\sf Y}}}
\def\Y{\mathsf{Y}}
\def\ssY{\mbox{\rm {\sf {\scriptsize Y}}}}
\def\Z{\mathsf{Z}}

\def\LCP{\mbox{\rm {\sf LCP}}}
\def\LPF{\mbox{\rm {\sf LPF}}}

\def\YZ{\mbox{\rm {\sf {\scriptsize Y$|$Z}}}}

\def\LZSCAN{\mbox{\sf LZscan}}
\def\EMLZSCAN{\mbox{\sf EM-LZscan}}

\begin{document}
\title{Lempel-Ziv Parsing in External Memory\thanks{This research is
    partially supported by Academy of Finland through grant 118653
    (ALGODAN) and grant 250345 (CoECGR).}}

\author{
Juha K{\"a}rkk{\"a}inen
\and
Dominik Kempa
\and
Simon J. Puglisi
}

\institute{
    Department of Computer Science,\\
    University of Helsinki\\
    Helsinki, Finland\\
    \email{\{firstname.lastname\}@cs.helsinki.fi}\\[1ex]
}

\date{}

\maketitle \thispagestyle{empty}

\begin{abstract}
  For decades, computing the LZ factorization (or LZ77 parsing) of a
  string has been a requisite and computationally intensive step in
  many diverse applications, including text indexing and data
  compression.  Many algorithms for LZ77 parsing have been discovered
  over the years; however, despite the
  increasing need to apply LZ77 to massive data sets, no algorithm to
  date scales to inputs that exceed the size of internal memory. In
  this paper we describe the first algorithm for computing the LZ77
  parsing in external memory. Our algorithm is fast in practice and
  will allow the next generation of text indexes to be realised for
  massive strings and string collections.
\end{abstract}

\section{Introduction}
\label{sec-intro}

For over three decades the Lempel-Ziv (LZ77) factorization~\cite{ZL77}
has been a fundamental tool for compressing
data~\cite{fm2010,RLZspire2010,GG10,kn2010} and for string processing
-- in particular for the efficient detection of
periodicities~\cite{bct2012,kk1999,kk2003,kbk2003}.  Recently the
factorization has become the basis for several compressed full-text
self-indexes~\cite{kn2011,ggp2011,ggknp2012}.  These indexes are
designed to support efficient storage and fast searching of massive,
highly repetitive data sets such as web documents, whole genome
collections, and versioned collections of source code and multi-author
documents, such as Wikipedia.

In order for these LZ77-based self-indexes to be constructed, whole
collection LZ77 factorizations need to be computed. However, to our
knowledge, all current LZ77 algorithms require large amounts of memory
and are essentially ``in-memory'' algorithms: they have poor locality
of memory reference, do not scale to external memory (disk), and so
are incapable of factorizing massive strings. In this paper we address
this shortcoming and design an LZ77 factorization algorithm capable of
scaling to data that substantially exceeds the size of RAM.

\paragraph{Our contribution.} 
We have designed and implemented the first external memory algorithm
for LZ77 factorization. It is based on the recent \LZSCAN\
algorithm~\cite{kkp2013-sea}, which itself is a space-efficient
algorithm, able compute the LZ factorization of a string of length $n$
using less than $2n$ bytes of memory. The new external memory version, 
\EMLZSCAN\, scales beyond the main memory size. Furthermore, our
experiments show that \EMLZSCAN\ is significantly faster than \LZSCAN\
already for much smaller files.

Theoretically, the time complexity of \EMLZSCAN\ is
$\Oh(n^2t_{\rank}/M)$, where $M$ is the size of the main memory and
$t_{\rank}$ is the time complexity of the rank operation on strings
(see, e.g.,~\cite{bgnn2010}).  The disk accesses are sequential with a
total I/O volume of $\Oh(n^2/M)$. The quadratic time complexity means
that the practical scalability of the algorithm is limited but still
beyond any previous algorithm. A couple of days on an ordinary desktop PC
with 4GiB of RAM is sufficient for computing the LZ factorization of
any 10GiB file or a 40GiB highly repetitive file, which is the
particularly relevant case for LZ77-based indexes.

\paragraph{Related work.}
A recent survey~\cite{aciksty2012} and some even more recent
papers~\cite{kkp2013,kkp2013-sea} outline
the many
algorithms for LZ77 factorization, all of which operate in RAM and
most of which make use of the suffix array (SA) and
longest-common-prefix (LCP) array, as intermediate data
structures. Many of the algorithms compute the {\em LPF tables}, which
contain all the longest previous factors, not just the ones needed for
the LZ parsing. 
Several LPF algorithms (\cite{cis2008} is a notably simple algorithm
in this category) compute the LPF tables in a single
left-to-right pass over the SA and LCP arrays, with sublinear (in fact
only $\Oh(\sqrt{n})$ words) extra memory. These algorithms would thus
seem to be candidates for external memory LZ77 factorization, however
it is the computation of SA and LCP arrays that forms the bottleneck.
There exists an external memory implementation for computing the SA
and LCP arrays~\cite{eSAIS}, but it is limited by the large disk space
requirement of $54n$ bytes (compared to less than $2n$ bytes for
\EMLZSCAN\ ). 

Another way to reduce working space is to use compressed suffix
arrays, augmented with some auxilliary data
structures~\cite{og2011,kn2010,kkp2013-sea}. However, the experiments
in~\cite{kkp2013-sea} show that such algorithms are inferior to
\LZSCAN\ . 

\section{Basic Notation and Algorithmic Machinery}
\label{sec-preliminaries}

\paragraph{Strings.}
Throughout we consider a string $\X = \X[1..n] = \X[1]\X[2]\ldots
\X[n]$ of $|\X| = n$ symbols drawn from the alphabet $[0..\sigma-1]$.
We assume $\X[n]$ is a special ``end of string'' symbol, \$, smaller than
all other symbols in the alphabet.
The reverse of $\X$ is denoted $\reverse{\X}$.  For $i=1,\ldots,n$ we
write $\X[i..n]$ to denote the {\em suffix} of $\X$ of length $n-i+1$,
that is $\X[i..n] = \X[i]\X[i+1]\ldots \X[n]$.  We will often refer to
suffix $\X[i..n]$ simply as ``suffix $i$''. Similarly, we write
$\X[1..i]$ to denote the {\em prefix} of $\X$ of length $i$.
$\X[i..j]$ is the {\em substring} $\X[i]\X[i+1]\ldots \X[j]$ of $\X$
that starts at position $i$ and ends at position $j$. By $\X[i..j)$ we
denote $\X[i..j-1]$.  If $j < i$ we
define $\X[i..j]$ to be the empty string, also denoted by
$\varepsilon$.

\paragraph{LZ77.}
Before defining the LZ77 factorization, we introduce the concept of a
{\em longest previous factor} (LPF).  The LPF at position $i$ in
string $\X$ is a pair $\LPF_{\X}[i]=(p_i,\ell_i)$ such that, $p_i < i$,
$\X[p_i..p_i+\ell_i) = \X[i..i+\ell_i)$, and $\ell_i$ is maximized.
In other words, $\X[i..i+\ell_i)$ is the longest
prefix of $\X[i..n]$ which also occurs at some position $p_i < i$ in
$\X$. 
Note also
that there may be more than one potential source (that is, $p_i$
value), and we do not care which one is used.

The LZ77 factorization (or LZ77 parsing) of a string $\X$ is then just
a greedy, left-to-right parsing of $\X$ into longest previous
factors. More precisely, if the $j$th LZ factor (or {\em phrase}) in
the parsing is to start at position $i$, then we output $(p_i,\ell_i)$
(to represent the $j$th phrase), and then the $(j+1)$th phrase starts
at position $i+\ell_i$. The exception is the case $\ell_i=0$, which
happens iff $\X[i]$ is the leftmost occurrence of a symbol in $\X$. In
this case we output $(\X[i],0)$ (to represent $\X[i..i]$) and the next
phrase starts at position $i+1$.  When $\ell_i > 0$, the substring
$\X[p_i..p_i+\ell_i)$ is called the {\em source} of phrase
$\X[i..i+\ell_i)$. We denote the number of phrases in the LZ77 parsing
of $\X$ by $z$.

\paragraph{Matching Statistics.}
Given two strings $\Y$ and $\Z$, the matching statistics of $\Y$
w.r.t.~$\Z$, denoted $\MS_{\YZ}$, is an array of $|\Y|$ pairs,
$(p_1,\ell_1)$, $(p_2,\ell_2)$, ..., $(p_{|\ssY|},\ell_{|\ssY|})$,
such that for all $i \in [1..|\Y|]$, $\Y[i..i+\ell_i) =
\Z[p_i..p_i+\ell_i)$ is the longest substring starting at position $i$
in $\Y$ that is also a substring of $\Z$. The observant reader will
note the resemblance to the LPF array. Indeed, if we replace
$\LPF_{\Y}$ with $\MS_{\YZ}$ in the computation of the LZ factorization
of $\Y$, the result is the relative LZ factorization of $\Y$
w.r.t.~$\Z$~\cite{RLZspire2010}.

\section{LZ77 Factorization in External Memory}
\label{sec-algorithm}

In this section we first describe the scanning-based, block-oriented  
LZ77 factorization algorithm called \LZSCAN\ that was introduced
in~\cite{kkp2013-sea}. We then present the modifications to \LZSCAN\
to make it run efficiently in external memory.

\subsection{Basic Algorithm}

Conceptually \LZSCAN\ divides $\X$ up into $d=\lceil n/b \rceil$ fixed size
blocks of length $b$: $\X[1..b]$, $\X[b+1..2b]$, ... . 
In the description that follows we will
refer to the block currently under consideration as $\B$, and to the
prefix of $\X$ that ends just before $\B$ as $\A$. Thus, if $\B =
\X[kb+1..(k+1)b]$, then $\A = \X[1..kb]$.  

To begin, we will assume no LZ factor or its source
crosses a boundary of the block $\B$. Later we will show how to remove
these assumptions.

The outline of the algorithm for processing a block $\B$ is shown below.
\begin{enumerate}
\item Compute $\MS_{\A|\B}$
\item Compute $\MS_{\B|\A}$ from $\MS_{\A|\B}$, $\SA_{\B}$ and $\LCP_{\B}$
\item Compute $\LPF_{\A\B}[kb+1..(k+1)b]$ from $\MS_{\B|\A}$ and
  $\LPF_{\B}$
\item Factorize $\B$ using $\LPF_{\A\B}[kb+1..(k+1)b]$
\end{enumerate}

\paragraph{Step 1: Computing Matching Statistics.}
Similarly to most algorithms for computing the matching statistics, we
first construct some data structures on $\B$ and then scan $\A$.  For
the details of the data structures we refer
to~\cite{kkp2013-sea}. The key properties are the space requirement of
$27b$ bytes and linear time construction.

The scanning of $\A$ is the computational bottleneck of the algorithm
in theory and practice. Theoretically, the time complexity of Step~1
is $\Oh((|\A|+|\B|)t_{\rank})$, where $t_{\rank}$ is the time complexity
of the rank operation on strings over the alphabet $\Sigma$ 
(see, e.g.,~\cite{bgnn2010}).
Thus the total time complexity of \LZSCAN\ is $\Oh(dnt_{\rank})$.  In
practice, each step of the scan may involve a substantial amount of
work in navigating the data structures. However, each character of $\A$
is accessed only once, and this is mostly done sequentially from right
to left.

An important optimization, called skipping trick, speeds up the
computation for highly repetitive inputs. It takes advantage of
repetition present in $\A$ that was found in the previous stages of
the algorithm. Consider an LZ factor $\A[i..i+\ell)$.  Because, by
definition, $\A[i..i+\ell)$ occurs earlier in $\A$ too, any source of
an LZ factor of $\B$ that is completely inside $\A[i..i+\ell)$ could
be replaced with an equivalent source in that earlier occurrence.
Thus such factors can be skipped during the computation of
$\MS_{\A|\B}$ without an effect on the factorization.

More precisely, if during the scan we compute $\MS_{\A|\B}[j]=(p,k)$
and find that $i \leq j < j + k \leq i+\ell$ for an LZ factor
$\A[i..i+\ell)$, we will compute $\MS_{\A|\B}[i-1]$ and continue the
scanning from $i-1$. However, we will do this only for long phrases
with $\ell\ge 40$.  To compute $\MS_{\A|\B}[i-1]$ from scratch, we use
right extension operations implemented by a binary search on
$\SA$. This is the only situation, where a part of $\A$ is scanned from
left to right, but still sequentially.

\paragraph{Step 2: Inverting Matching Statistics.}

With the help of $\SA_{\B}$ and $\LCP_{\B}$, we can \emph{invert}
$\MS_{\A|\B}$ to obtain $\MS_{\B|\A}$, which is what we need for LZ77
factorization. Again, we refer to~\cite{kkp2013-sea} for details of
the inversion algorithm and give only the key properties.  The
algorithm accesses each entry of $\MS_{\A|\B}$ (except those skipped
by the skipping trick) once, in an arbitrary order, and processes the
entry in constant time. Thus we do not need to store $\MS_{\A|\B}$
but can process each entry as soon as it is produced in Step~1.
The rest of the computation takes $\Oh(b)$ time.

\paragraph{Step3: Computing LPF.}

Consider the pair $(p,\ell)=\LPF_{\A\B}[i]$ for $i\in[kb+1..(k+1)b]$
that we want to compute and assume $\ell>0$ (otherwise $i$ is the
position of the leftmost occurrence of $\X[i]$ in $\X$, which we can
easily detect).  Clearly, either $p\le kb$ and
$\LPF_{\A\B}[i]=\MS_{\B|\A}[i]$, or $kb < p < i$ and
$\LPF_{\A\B}[i]=(kb+p_{\B},\ell_{\B})$, where
$(p_{\B},\ell_{\B})=\LPF_{\B}[i-kb]$.  Thus computing $\LPF_{\A\B}$
from $\MS_{\B|\A}[i]$ and $\LPF_{\B}$ is easy.

The above is true if the sources do not cross the block
boundary, but the case where $p\le kb$ but $p+\ell > kb+1$ is not
handled correctly. An easy correction is to replace $\MS_{\A|\B}$ with
$\MS_{\A\B|\B}[1..kb]$ in all of the steps. This does not affect the
essential features of the algorithm.

\paragraph{Step 4: Parsing.}

We use the standard LZ77 parsing to factorize $\B$ except
$\LPF_{\B}$ is replaced with $\LPF_{\A\B}[kb+1..(k+1)b]$.

So far we have assumed that every block starts with a new phrase, or,
put another way, that a phrase ends at the end of every block. Let
$\Z=\X[i..(k+1)b]$ be the last factor in $\B$ after we have factorized
$\B$ as described above. This may not be a true LZ factor when
considering the whole $\X$ because the true LZ factor may continue
beyond the end of $\B$.  If $|\Z| \le b/2$, we start the next block at
$i$ instead of $(k+1)b+1$, and compute the true phrase starting at $i$
while processing that block. This at most doubles the computation.  If
$|\Z| > b/2$, we need to do something more sophisticated.
In~\cite{kkp2013-sea}, a modified constant extra space pattern
matching algorithm by Crochemore~\cite{c1992} is used for finding the
true phrase.

\subsection{Implementation in External Memory}
\label{sec-ms}

Next we describe an external memory adaptation of \LZSCAN\ called $\EMLZSCAN$.

\paragraph{Block data structures.}

The structures constructed for the current block $\B$ are essentially
the same in \EMLZSCAN\ as in \LZSCAN\ and are kept in memory during the
processing of $\B$. There are two notable differences: $\B$ itself is
read from disk and held in memory during the stage, and we replace
32-bit integers with 40-bit integers to represent positions in the
whole text (but still 32-bit integers for positions in $\B$).  These
changes raise the peak memory usage of the data structures from $27b$
bytes to $29b$ bytes. We have also implemented a 32-bit version of
\EMLZSCAN\ that needs $28n$ bytes of space and runs slightly faster
because processing 40-bit integers incurs a small overhead.

\paragraph{Scanning $\A$.} 

The scanning of $\A$ is performed by reading $\A$ from disk into a
buffer of size 256KiB. We store $\X$ in reverse order on disk so that
the backward scan of $\A$ involves reading in forward direction. This
seems to make the algorithm faster even when the time for reversing
$\X$ is included.

To implement the skipping trick, we need to identify phrases of length
40 or more during the scan. They are stored in a separate file in
reverse sequential order, which is then scanned in synchrony with
$\A$. Since the file grows backwards during the computation, we create
a file of the maximum size, which is $n/5$ bytes, in the beginning,
and fill it starting from the end.

\paragraph{Long incomplete phrases.}

A potentially incomplete phrase at the end of a block is handled the
same way as in $\LZSCAN$. If the incomplete phrase is short, we start
the next block at the beginning of the phrase, and if it is long, we
use the modified pattern matching algorithm by Crochemore to compute
the full phrase. Being a constant extra space algorithm, Crochmore's
algorithm works in the external memory setting as long as the full
phrase fits in memory. Even longer phrases can be handled as follows.
Find the occurrences of the longest prefix of the phrase that fits in
memory and write the end positions of the occurrences to disk. Then
read the next part of the phrase and find its occurrences that start
at those end positions and so on. Crochmore's algorithm can be
modified to handle this too~\cite{kkp2013-psc}. Note that this does not change the
complexity of the algorithm since each additional round of computation
advances the factorization by $\Omega(b)$ steps.

\paragraph{Complexity.}

The CPU time complexity of \EMLZSCAN\ is the same as \LZSCAN\,
$\Oh(dnt_{\rank})$, where $d=\Oh(n/M)$ and $M$ is the size of the main
memory. Thus the time complexity is $\Oh(n^2t_{\rank}/M)$.  The I/O
complexity is dominated by the scans of $\A$. Thus the total I/O
volume is $\Oh(n^2/M)$. 

\section{Experimental Results} \label{sec:experiments}

\begin{table*}[bt]
  \centering {
  \begin{tabular}[tab:space-basic]{l@{\hspace{2.3em}}l@{\hspace{2.3em}}r@{\hspace{2.3em}}r@{\hspace{2.3em}}l@{\hspace{2.3em}}l}
\hline
Name      & $n$      & $\sigma$ & $n/z$ & Description \\
\hline
hg        & 5.85 GiB & 31       & 18.98 & 2 $\times$ Human genome \\
enwik     & 8 GiB    & 209      & 20.45 & English Wikipedia XML \\
countries & 40.5 GiB & 203      & 3185  & Wikipedia version database \\
cere      & 31 GiB   & 5        & 4849  & Yeast DNA \\
\hline
  \end{tabular} }\vspace{1ex}
\caption[lof]{Files used in the experiments.
  The value of $n/z$ (the average length of a phrase in the LZ factorization) is
  included as a measure of repetitiveness.}
  \label{tab-files}
\end{table*}

We implemented two versions of the algorithm described in this paper: the first
(32-bit) can parse files up to 4 GiB, the second (40-bit) is capable of handling
texts up to 1 TiB. We simulate 40-bit integers as pairs of 32- and 8-bit integers.
This slightly deteriorates the speed but compared to 32-bit version increases
the space usage only by $b$ bytes. The implementations will be made available at
\url{http://www.cs.helsinki.fi/group/pads/}

\footnotetext[1]{\url{http://hgdownload.soe.ucsc.edu/goldenPath/hg19/chromosomes/}}
\footnotetext[2]{\url{ftp://public.genomics.org.cn/BGI/yanhuang/fa/}}
\footnotetext[3]{\url{http://dumps.wikimedia.org/enwiki/}}
\footnotetext[4]{\url{http://en.wikipedia.org/wiki/List_of_countries_by_GDP_(nominal)}}
\footnotetext[5]{\url{http://pizzachili.dcc.uchile.cl/repcorpus.html}}

\paragraph{Data set.}
In our experiments we used the following files:\newline
- hg: a concatenation of two different Human genomes
\footnotemark[1]${}^{,}$\footnotemark[2],\newline
- enwik: a prefix of the latest (20130403) English Wikipedia dump\footnotemark[3],\newline
- countries: a concatenation of all versions (as of April 16, 2013) of Wikipedia
articles about 40 large countries\footnotemark[4],\newline
- cere: a concatenation of multiple copies of \textit{cere} testfile from
Pizza\&Chili repetitive corpus\footnotemark[5], each randomly mutated with respect
to original with rate 0.01\%.

\begin{figure}
\minipage{0.50\textwidth}
  \includegraphics[trim = 0mm 25mm 0mm 0mm, width=\linewidth]{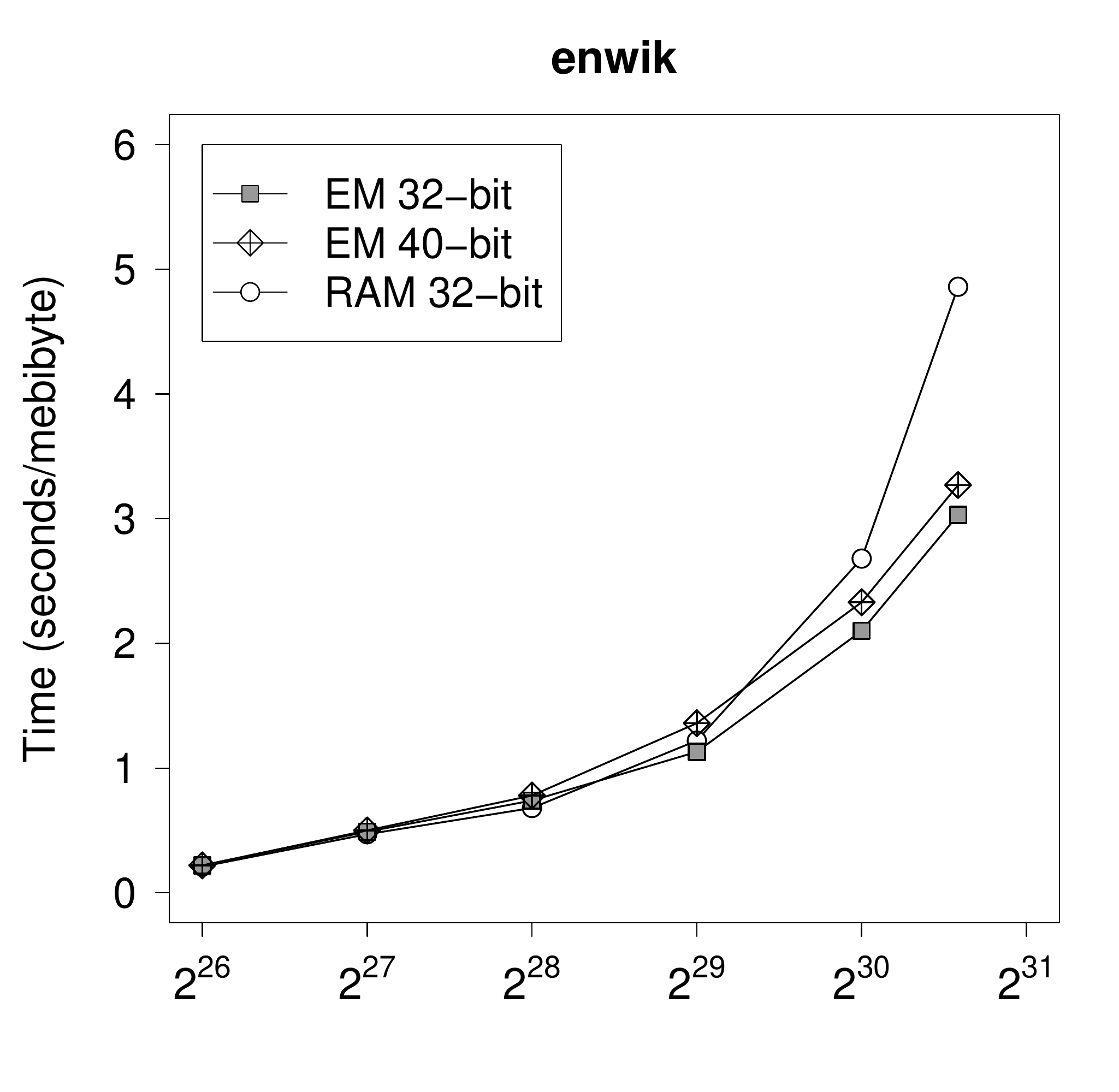}
\endminipage\hfill
\minipage{0.50\textwidth}
  \includegraphics[trim = 20mm 25mm -20mm 0mm, width=\linewidth]{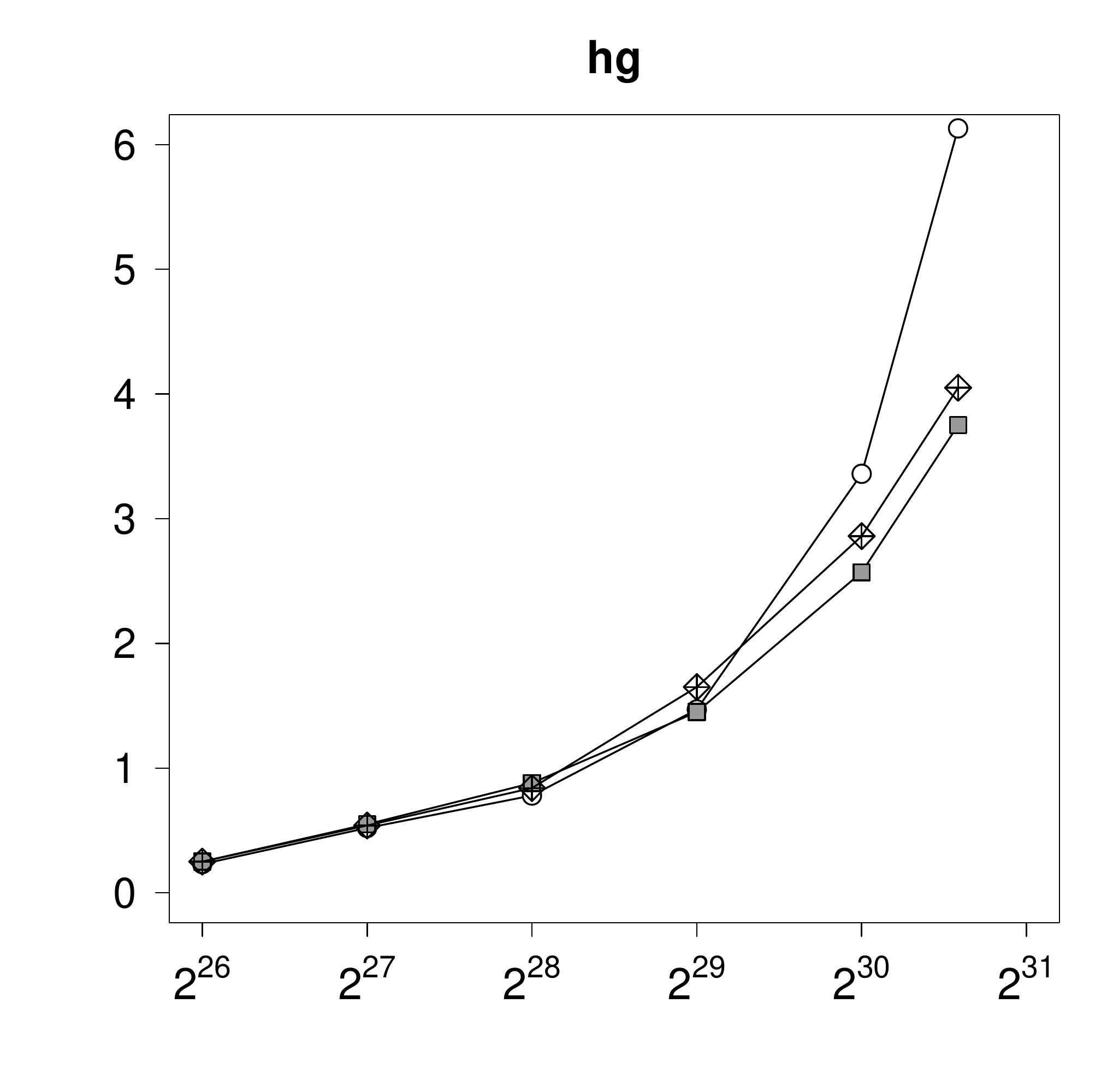}
\endminipage
\vspace{1ex}
\newline
\minipage{0.50\textwidth}
  \includegraphics[trim = 0mm 25mm 0mm 0mm, width=\linewidth]{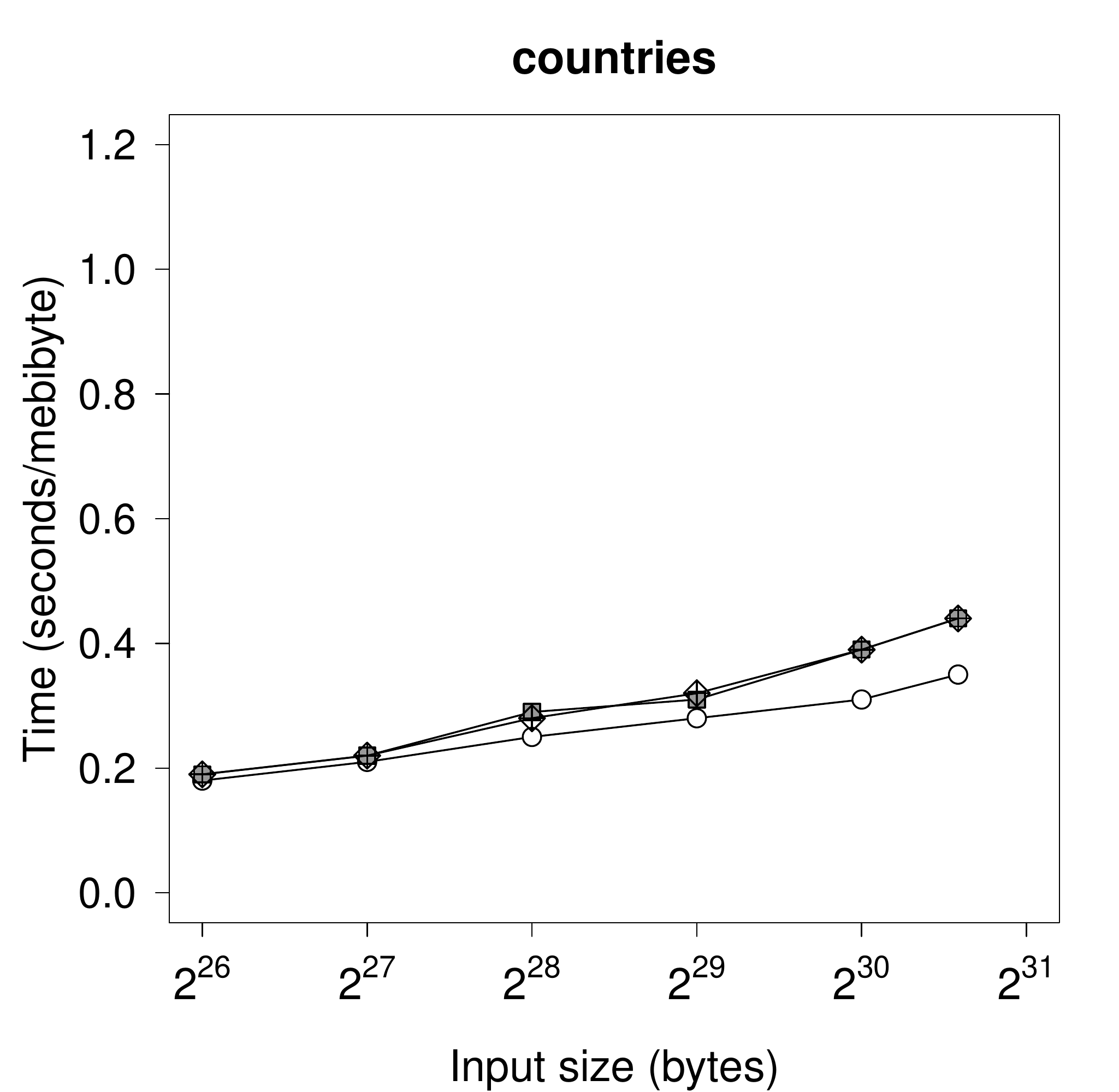}
\endminipage\hfill
\minipage{0.50\textwidth}
  \includegraphics[trim = 20mm 25mm -20mm 0mm, width=\linewidth]{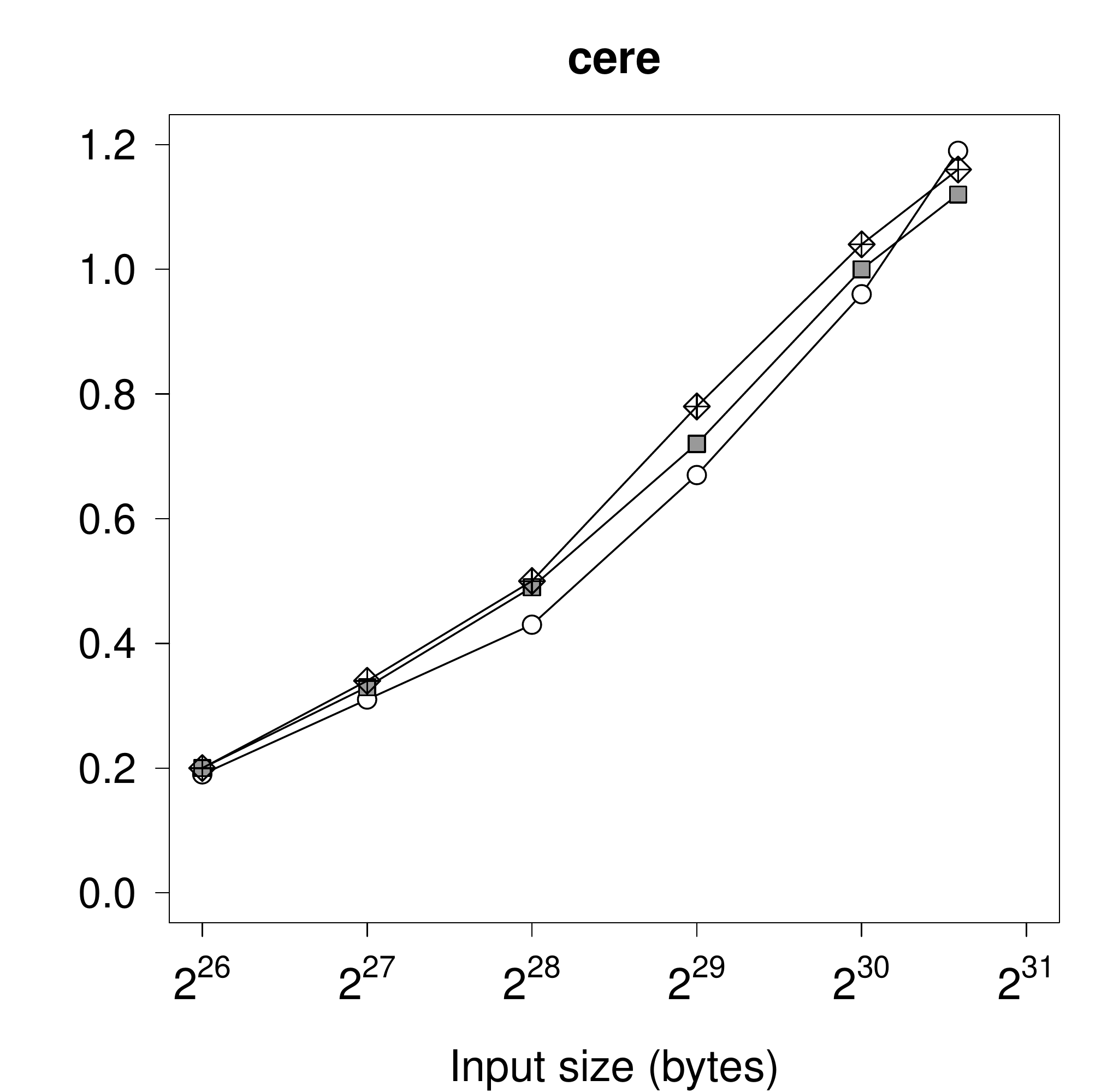}
\endminipage
\vspace{5ex}
\caption{Comparison of EM and RAM variants of $\LZSCAN$.}
\label{fig-comparison}
\end{figure}

Statistics about the files are summarized in Table~\ref{tab-files}.

\paragraph{Setup.}
We performed experiments on a set of three identical machines, each equipped with 
a 3.16GHz Intel Core 2 Duo CPU with 6144KiB L2 cache and 4GiB of main memory. Each
machine had two 320GiB hard drives, one held the input text used during experiments
and the other stored the operating system as well as all auxiliary files created by algorithms. The machines
had no other significant CPU tasks running and only a single thread of execution was used. 
The OS was Linux (Ubuntu 12.04, 64bit) running kernel 3.2.0. All programs were compiled using
{\tt g++} version 4.6.4 with {\tt-O3} {\tt-static} {\tt-DNDEBUG}
options. 
All reported runtimes are wallclock (real) times, recorded with the Unix {\tt gettimeofday} function.

\begin{figure}
\minipage{0.50\textwidth}
  \includegraphics[trim = 0mm 25mm 5mm 0mm, width=\linewidth]{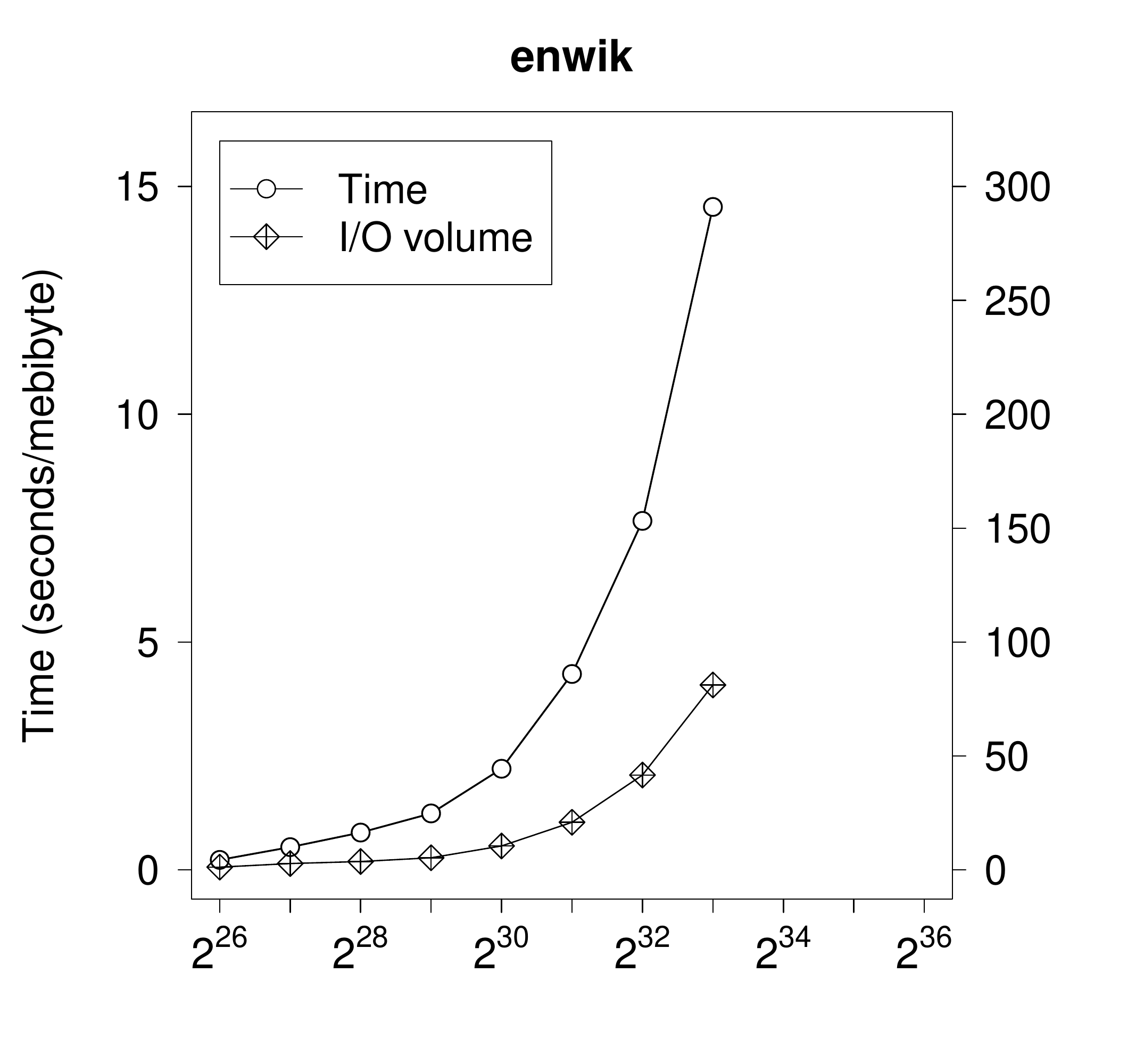}
\endminipage\hfill
\minipage{0.50\textwidth}
  \includegraphics[trim = 25mm 25mm -20mm 0mm, width=\linewidth]{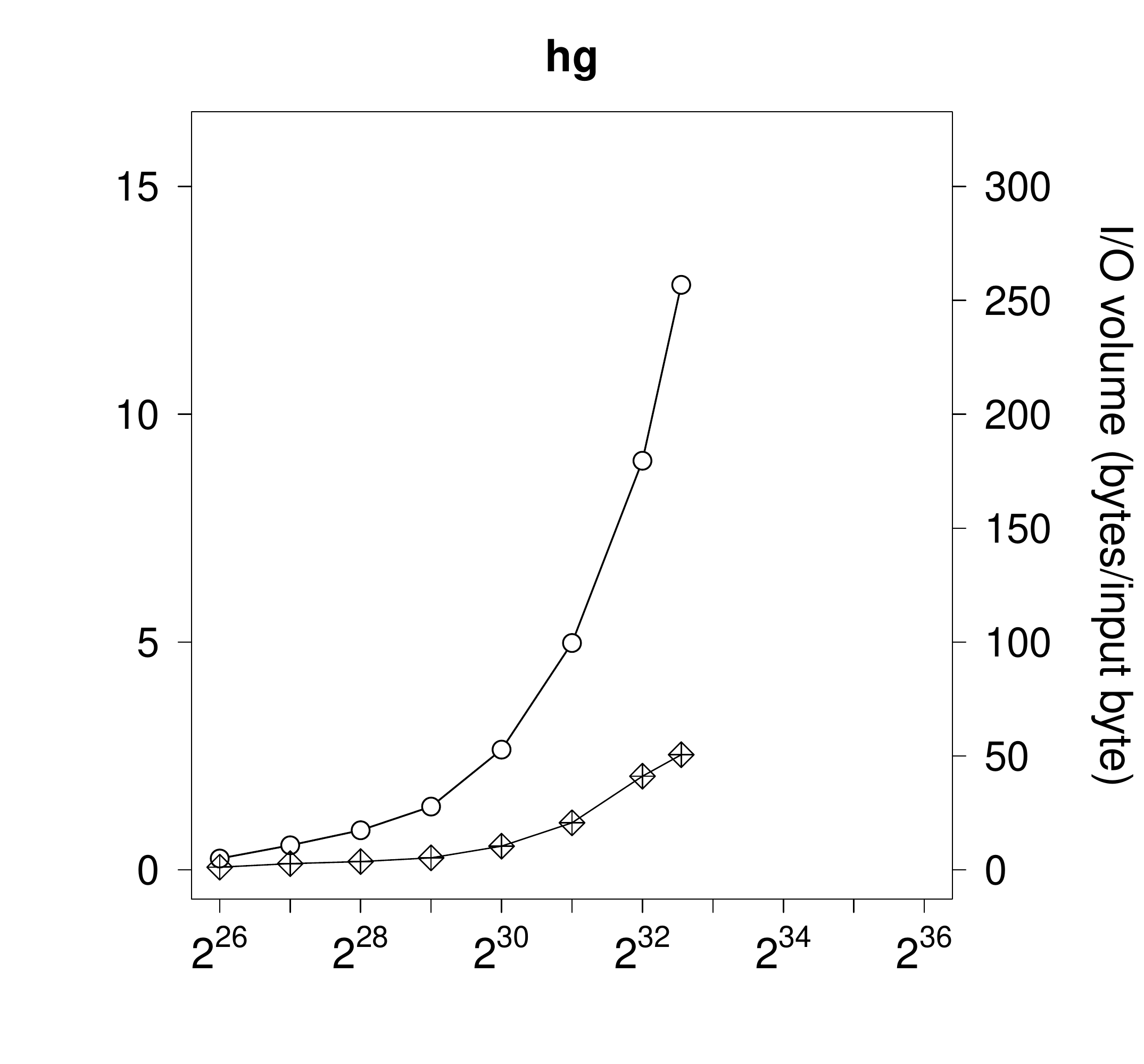}
\endminipage
\vspace{1ex}
\newline
\minipage{0.50\textwidth}
  \includegraphics[trim = 0mm 25mm 5mm 0mm, width=\linewidth]{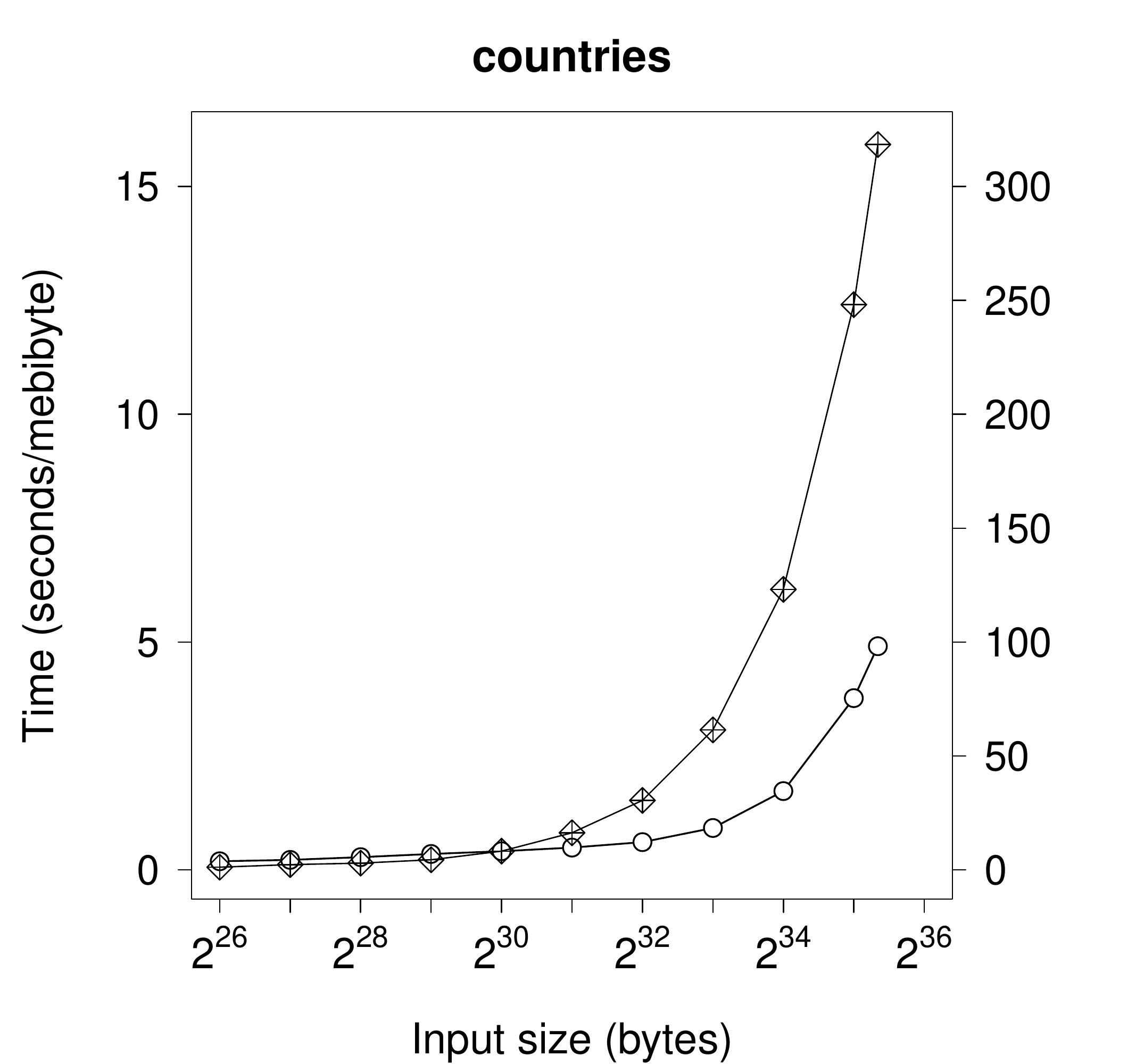}
\endminipage\hfill
\minipage{0.50\textwidth}
  \includegraphics[trim = 25mm 25mm -20mm 0mm, width=\linewidth]{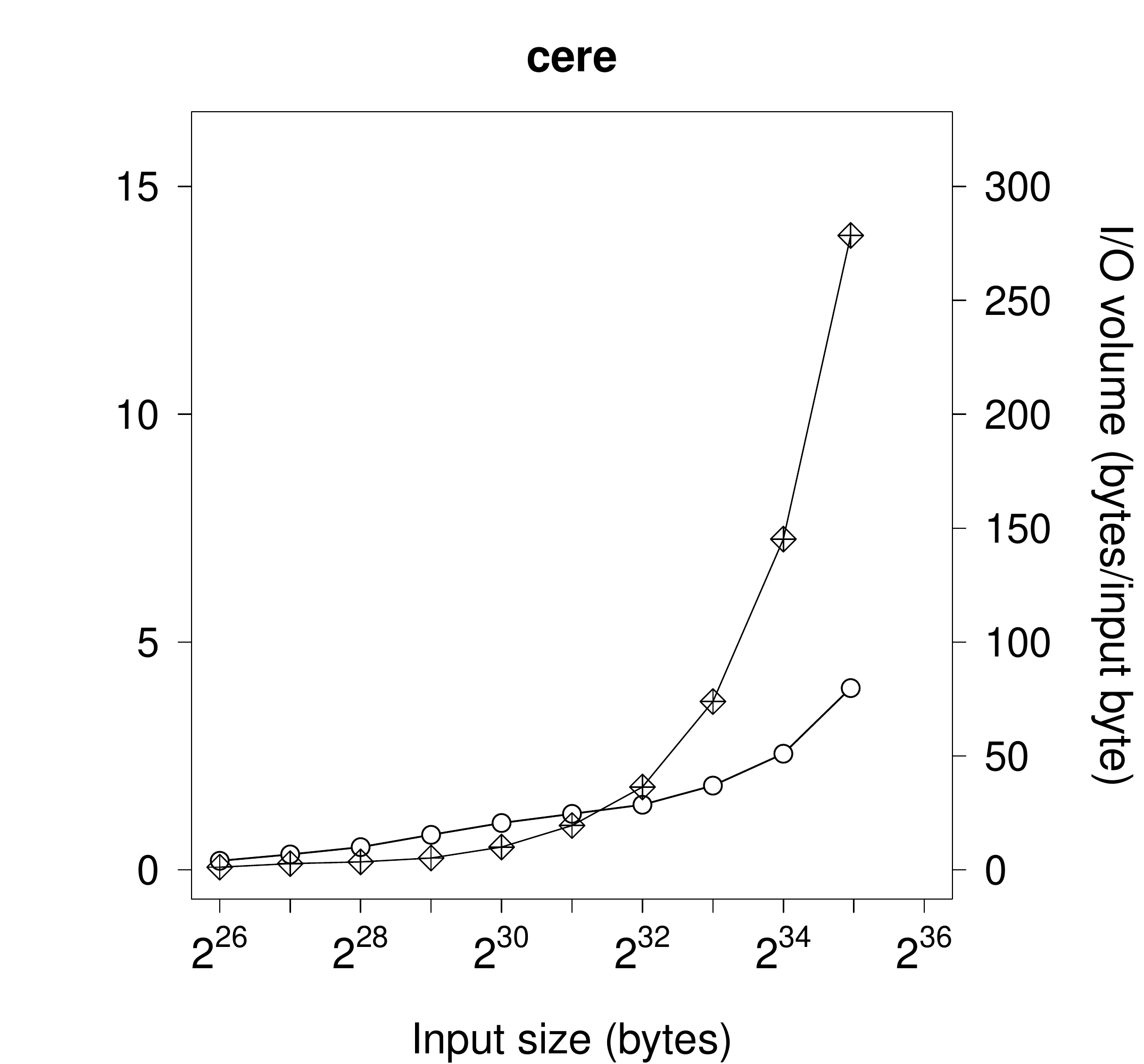}
\endminipage
\vspace{5ex}
\caption{Scalability of 40-bit $\EMLZSCAN$.}
\label{fig-scalability}
\end{figure}

\paragraph{Description.}
Our first experiment compares the speed of two $\EMLZSCAN$ versions to the in-memory
variant of $\LZSCAN$, described in~\cite{kkp2013-sea}. The parsing was computed for
increasing length prefixes of testfiles. All algorithms were limited to use 3GiB
of RAM in all runs. The results are given in Fig.~\ref{fig-comparison}.

The second experiment measures the scalability of $\EMLZSCAN$.
Similarly to previous experiment we perform the computation for various length
prefixes of input files.
The results are presented in Fig.~\ref{fig-scalability}. For each run we report
the runtime (scaled to prefix length) and the I/O volume, that is, the total
number of bytes transferred between RAM and disk normalized to bytes per text
symbol (here also byte).

\paragraph{Discussion.}

The comparison of EM and RAM $\LZSCAN$ clearly exhibits dependence on the 
repetitiveness of the input. For non-repetitive files, where the runtime is
dominated by the CPU computation, the EM algorithms are faster due to the use of
larger blocks (no space is necessary to hold the input text, as in the RAM
version). Bigger blocks imply less streaming phrases during which the matching
statistics computation is performed (which is the bottleneck in the parsing of
non-repetitive files -- the streaming speed does not exceed 3 MiB/s).

On repetitive inputs the RAM version is slightly faster because of
the streaming speed, which (mostly due to the skip trick) exceeds 200 MiB/s,
whereas the disk latency is limiting the streaming speed of EM variants to
$\sim$70MiB/s.

In all cases the 32-bit version of $\EMLZSCAN$ is slightly faster than the 40-bit
version. This is caused by the memory layout of two arrays holding 40-bit integers.
Accessing each integer requires reaching two distant memory locations possibly attracting
two cache misses, unlike the 32-bit version where at most one cache miss can occur.

As observed from Figure~\ref{fig-scalability}, the 40-bit $\EMLZSCAN$ scales really
well for highly repetitive inputs (which are the most common targets of LZ77
factorization), e.g. parsing of \textit{countries} (40.5 GiB) file took 2.3 days
on our commodity hardware. The usability of $\EMLZSCAN$ on non-repetitive inputs
cannot extend much beyond 2 to 3 times the size of RAM due to its quadratic complexity, but in
practice non-repetitive inputs are less often treated with LZ77 because of its
limited compression for such texts.

\bibliographystyle{splncs03}
\bibliography{lz}

\end{document}